\documentclass[aps,prb,twocolumn,showpacs,superscriptaddress,floatfix]{revtex4-1}
\usepackage{amsmath,amssymb}
\usepackage{graphicx}
\usepackage{epsfig}
\usepackage{color}
\usepackage{rotating}
\usepackage{bm}
\usepackage{setspace}

\begin{document}
\title{Electronic structure of the ingredient planes of cuprate superconductor Bi$_2$Sr$_2$CuO$_{6+\delta}$: a comparison study with Bi$_2$Sr$_2$CaCu$_2$O$_{8+\delta}$}
\author{Yan-Feng Lv}
\author{Wen-Lin Wang}
\author{Hao Ding}
\author{Yang Wang}
\affiliation{State Key Laboratory of Low-Dimensional Quantum Physics, Department of Physics, Tsinghua University, Beijing 100084, China}
\author{Ying Ding}
\affiliation{National Lab for Superconductivity, Beijing National Laboratory for Condensed Matter Physics, Institute of Physics, Chinese Academy of Sciences, Beijing, 100190, China}
\author{Ruidan Zhong}
\author{John Schneeloch}
\author{G. D. Gu}
\affiliation{Condensed Matter Physics and Materials Science Department, Brookhaven National Laboratory, Upton, New York 11973, USA}
\author{Lili Wang}
\author{Ke He}
\author{Shuai-Hua Ji}
\affiliation{State Key Laboratory of Low-Dimensional Quantum Physics, Department of Physics, Tsinghua University, Beijing 100084, China}
\affiliation{Collaborative Innovation Center of Quantum Matter, Beijing 100084, China}
\author{Lin Zhao}
\affiliation{National Lab for Superconductivity, Beijing National Laboratory for Condensed Matter Physics, Institute of Physics, Chinese Academy of Sciences, Beijing, 100190, China}
\author{Xing-Jiang Zhou}
\affiliation{National Lab for Superconductivity, Beijing National Laboratory for Condensed Matter Physics, Institute of Physics, Chinese Academy of Sciences, Beijing, 100190, China}
\affiliation{Collaborative Innovation Center of Quantum Matter, Beijing 100084, China}
\author{Can-Li Song}
\email[]{clsong07@mail.tsinghua.edu.cn}
\author{Xu-Cun Ma}
\email[]{xucunma@mail.tsinghua.edu.cn}
\author{Qi-Kun Xue}
\email[]{qkxue@mail.tsinghua.edu.cn}
\affiliation{State Key Laboratory of Low-Dimensional Quantum Physics, Department of Physics, Tsinghua University, Beijing 100084, China}
\affiliation{Collaborative Innovation Center of Quantum Matter, Beijing 100084, China}
\date{\today}

\begin{abstract}
By means of low-temperature scanning tunneling microscopy, we report on the electronic structures of BiO and SrO planes of Bi$_2$Sr$_2$CuO$_{6+\delta}$ (Bi-2201) superconductor prepared by argon-ion bombardment and annealing. Depending on post annealing conditions, the BiO planes exhibit either a pseudogap (PG) with sharp coherence peaks and an anomalously large gap magnitude of 49 meV or van Hove singularity (VHS) near the Fermi level, while the SrO is always characteristic of a PG-like feature. This contrasts with Bi$_2$Sr$_2$CaCu$_2$O$_{8+\delta}$ (Bi-2212) superconductor where VHS occurs solely on the SrO plane. We disclose the interstitial oxygen dopants ($\delta$ in the formulas) as a primary cause for the occurrence of VHS, which are located dominantly around the BiO and SrO planes, respectively, in  Bi-2201 and Bi-2212. This is supported by the contrasting structural buckling amplitude of BiO and SrO planes in the two superconductors. Our findings provide solid evidence for the irrelevance of PG to the superconductivity in the two superconductors, as well as insights into why Bi-2212 can achieve a higher superconducting transition temperature than Bi-2201, and by implication, the mechanism of cuprate superconductivity.
\end{abstract}
\maketitle
\begin{spacing}{0.99}
In high-transition temperature ($T_\textrm{c}$) cuprate superconductors, the maximum $T_\textrm{c}$ ($T_\textrm{c, max}$) varies substantially with the number ($n$) of CuO$_2$ planes in one unit cell, and peaks at $n$ = 3 \cite{Eisaki2004effect}. In bismuth-based cuprates, for example, $T_\textrm{c, max}$ is approximately 34 K, 90 K, and 110 K for $n$ = 1, 2, 3, respectively \cite{Feng2002electronic}. It has led to numerous competing proposals to explain this intriguing phenomenon, which include interlayer quantum tunneling of Cooper pairs \cite{chakravarty1993interlayer, chakravarty2004explanation, Nishiguchi2013superconductivity}, intralayer hopping \cite{Pavarini2001band}, the energy level of apical oxygen \cite{Ohta1991Apex}, magnetic exchange interactions \cite{Dean2014itinerant, Ellis2015correlation}, and so on. Thus far, however, a consensus on which factor controls $T_\textrm{c}$ in cuprate superconductors has not yet been reached, partially due to a lack of knowledge about the detailed electronic properties outside the superconducting CuO$_2$ planes, which are anti-ferromagnetic insulators without doping. Indeed, it has long been hypothesized that out-of-plane apical oxygen plays a primary role in determining the optimal $T_\textrm{c}$ of cuprate superconductors \cite{Ohta1991Apex, Chen1992out, Yin2009tunning}. Identification of the out-of-plane effects are thus imperative to understanding $T_\textrm{c, max}$ and superconductivity mechanism in cuprates \cite{slezak2008imaging}, but extremely challenging because technically it demands nonstandard, profile-based preparation and imaging techniques.

Cuprate superconductivity develops when the insulating CuO$_2$ planes are either electron or hole-doped by substitutional or interstitial chemical doping, e.g.\ excess oxygen dopants in the hole-doped cuprate superconductors. In addition to enabling superconductivity, the doping can lead to startling nanoscale electronic inhomogeneity and disordering \cite{pan2001microscopic}. The latter is usually detrimental to superconductivity \cite{Fujita2005effect}. However, its effect has been overemphasized over the past two decades \cite{lang2002imaging, gomes2007visualizing, Crocker2011NMR}: the experimental efforts in minimizing this secondary effect have demonstrated an enhancement in $T_\textrm{c}$ by only several Kelvins in bismuth-based cuprate superconductors \cite{Eisaki2004effect, Hobou2009enhancement}. In order to understand how the dopants boost superconductivity, scanning tunneling microscopy/spectroscopy (STM/STS) studies have been conducted to visualize the interstitial excess oxygen dopants in Bi$_2$Sr$_2$CaCu$_2$O$_{8+\delta}$ (Bi-2212) superconductor \cite{mcelroy2005atomic, zeljkovic2012imaging, zeljkovic2014nanoscale}. However, the central issue of how the excess oxygen dopants affect the electronic structure in the out-of-plane direction and thus that of CuO$_2$ planes remains unknown \cite{mcelroy2005atomic, zeljkovic2012imaging, zeljkovic2014nanoscale, Zhou2007correlating}.

Our recent argon-ion (Ar$^+$) bombardment and annealing (IBA) approach has enabled a layer-by-layer mapping by STM/STS of the out-of-plane electronic structures in Bi-2212 cuprate superconductor \cite{lv2015mapping}, and is particularly suited to addressing the above issues. Here we extend this technique to Bi$_2$Sr$_2$CuO$_{6+\delta}$ (Bi-2201) cuprate superconductor, which exhibits a lower $T_\textrm{c, max}$ but simpler crystal structure than Bi-2212. Our experiments were carried out in a Unisoku ultrahigh vacuum (UHV) cryogenic STM system with an ozone-assisted molecular beam epitaxy (MBE) chamber, in which an Ar$^+$ ion gun is installed, as detailed elsewhere \cite{lv2015mapping}. High quality Pb-doped Bi-2201 single crystals in the extremely over-doped region ($T_\textrm{c}$ = 4 K) were synthesized by a traveling solvent floating zone method \cite{lin2010high}, and used throughout. After IBA and post annealing under an ozone flux beam, the Bi-2201 samples were immediately inserted into the STM head for STM/STS measurements at 4.2 K. Polycrystalline PtIr tips were cleaned by $e$-beam heating in UHV and calibrated on MBE-grown Ag/Si(111) films. All STM topographies were acquired in a constant-current mode with a bias $V$ applied to the sample. Tunneling spectra were measured using a standard lock-in technique with a bias modulation of 2 meV at 931 Hz.
\end{spacing}

\begin{figure}[h]
\includegraphics[width=\columnwidth]{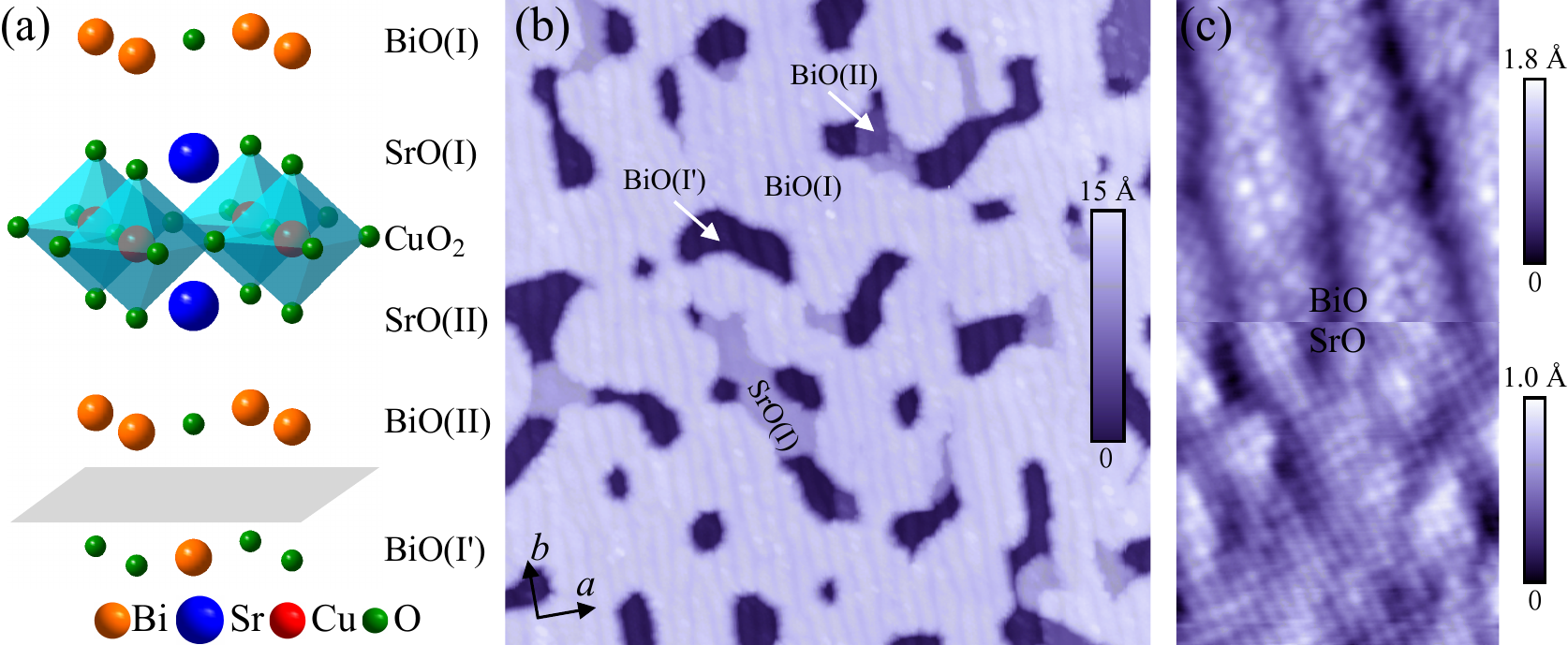}
\caption{(color online) (a) Schematic crystal structure of Bi-2201, with the gray parallelogram indicating the easily cleaved BiO planes. The CuO$_6$ octahedra is cyan-colored. (b) STM topography (100 nm $\times$ 100 nm, $V$ = 1.5 V, $I$ = 30 pA) of Ar$^+$ bombarded and annealed Bi-2201 cuprate superconductor, exposing four constituent planes of BiO(I), SrO(I), BiO(II) and BiO(I$^\prime$). $a$ and $b$ correspond to the in-plane crystallographic axes ($Cmmm$ space group), with $a=b=$ 5.4 {\AA}. (c) Zoom-in (6 nm $\times$ 6 nm, $V$ = 0.1 V, $I$ = 30 pA) on the exposed BiO and SrO planes.
}
\end{figure}

Drawn in Fig.\ 1(a) is the crystallographic structure of Bi-2201, which consists of two building blocks (BiO and SrO) other than one superconducting CuO$_2$ plane. The strong hybridization between the $p_z$ orbital of apical oxygen in the two adjacent SrO blocks and out-of-plane Cu $d_{3r^2-z^2}$ orbital leads to the formation of CuO$_6$ octahedra, analogous to the CuO$_5$ pyramid in Bi-2212. Figure 1(b) depicts a constant-current topographic image of Bi-2201, which has been sputtered by 500 eV Ar$^+$ under a pressure of 1 $\times$ 10$^{-5}$ Torr and followed by UHV annealing at 500$^\textrm{o}$C. Four constituent planes of BiO(I), SrO(I), BiO(II) and BiO(I$^\prime$), defined in Fig. 1(a), are discernible. The failure to obtain CuO$_2$ planes might most probably originate from the fact that CuO$_2$ couples strongly with a pair of SrO layers in Bi-2201, unlike only a SrO layer in Bi-2212 \cite{lv2015mapping}. A detailed examination reveals an unreconstructed and atomically flat surface on all the exposed planes, as shown in the zoom-in images in Fig.\ 1(c), suggesting that the electronic spectra measured on the exposed planes are characteristic of their bulk counterparts in Bi-2201. Remarkably the incommensurate structural buckling, namely the $b$-axis supermodulation, appears significantly weaker on the SrO plane than that on BiO, as will be discussed in detail later.

\begin{figure*}[t]
\includegraphics[width=1.73\columnwidth]{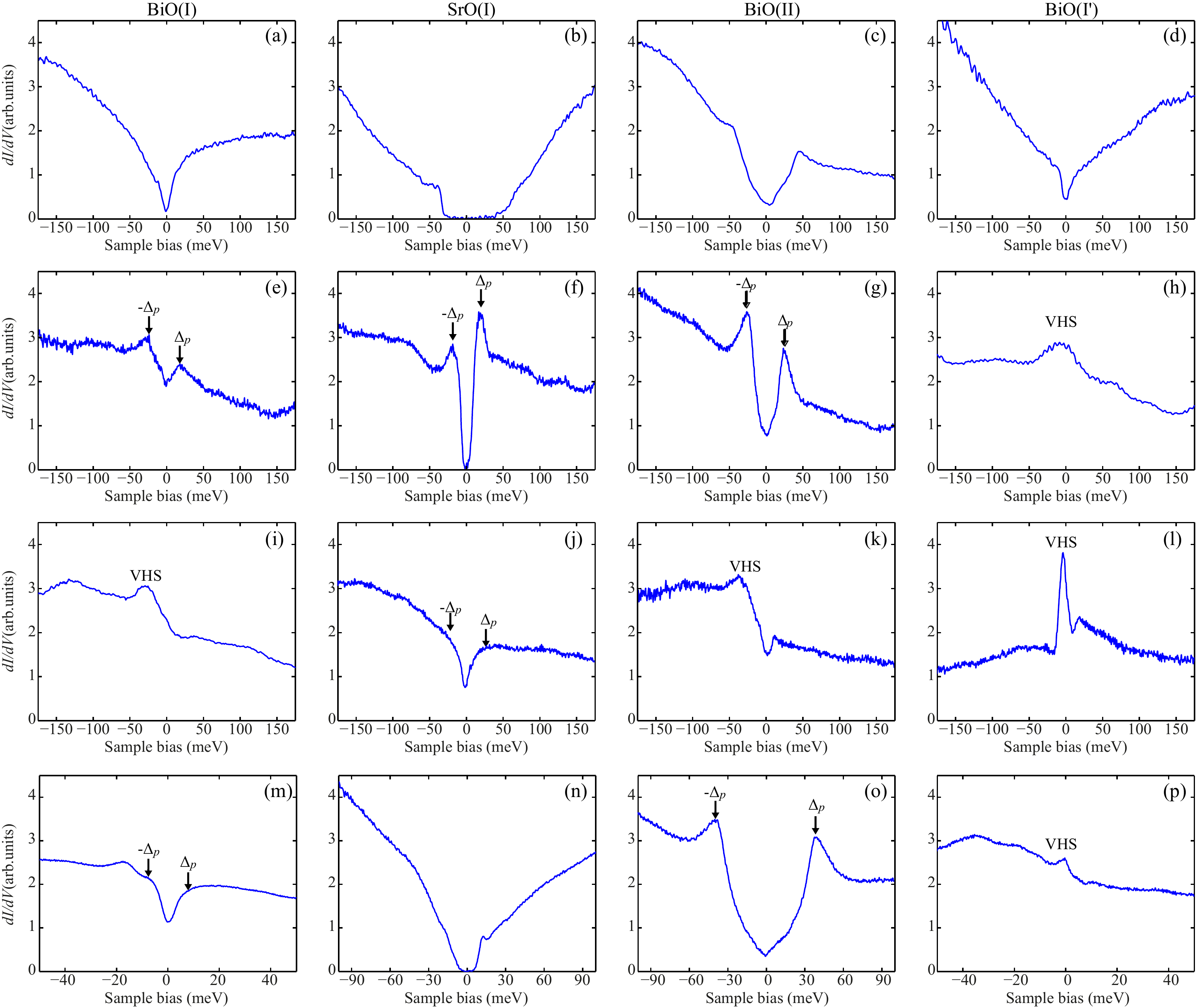}
\caption{(color online) Representative differential conductance $dI/dV$ spectra on the exposed BiO(I), SrO(I), BiO(II) and BiO(I$^\prime$) constituent planes of Ar$^+$ bombarded Bi-2201 single crystal, followed by a sequence of (a-d) UHV annealing at 500$^\textrm{o}$C, (e-h) ozone exposure of 18000 Langmuir, (i-l) additional ozone exposure of 18000 Langmuir, and (m-p) UHV annealing at 400$^\textrm{o}$C. The ozone exposure was performed at an optimal sample temperature of 450$^\textrm{o}$C. Black arrows denote the energy positions of PG. The VHS close to $E_F$ is found on BiO. Setpoint: (a-l) $V$ = 0.2 V, $I$ = 200 pA; (m, p) $V$ = 0.05 V, $I$ = 250 pA; (n, o) $V$ = 0.1 V, $I$ = 250 pA.
}
\end{figure*}

Figure 2 summarizes the electronic spectra on the varying planes of BiO(I), SrO(I), BiO(II) and BiO(I$^\prime$) in Ar$^+$ bombarded Bi-2201, followed by UHV annealing to restore the flat surface [Figs.\ 2(a-d)], post oxidation annealing under an ozone flux beam of 1.0 $\times$ 10$^{-5}$ Torr to add interstitial oxygen [Figs.\ 2(e-l)], and further moderate reduction annealing under UHV condition to remove a small amount of interstitial oxygen dopants [Figs.\ 2(m-p)]. The as-sputtered plus UHV annealed samples exhibit $dI/dV$ spectra with a substantial depletion in density of states (DOS) near the Fermi level ($E_F$) on the SrO plane and a pseudogap (PG, or DOS dip)-like feature on BiO [Figs.\ 2(a-d)], respectively. Such characteristics, primarily owing to a substantial loss of near-surface interstitial oxygen during IBA, resemble closely with those previously reported in Bi-2212 \cite{lv2015mapping}. A subsequent oxidation annealing can put the interstitial oxygen dopants back and recover the superconductivity of Bi-2201 studied [Figs.\ 2(e-p)], judged by comparing the $dI/dV$ spectra of BiO(I) plane with ``standard'' ones of the freshly cleaved superconducting Bi-2201 samples \cite{piriou2011first, Kugler2001Scanning, boyer2007imaging, he2014fermi}. A closer inspection of these spectra under different annealing conditions has revealed three fundamental findings regarding cuprate superconductors, which we discuss in turn below.

Firstly, a pseudogap occurs not only on the BiO planes in a wide variety of situations [Figs.\ 2(e), 2(g), 2(m) and 2(o)], but also on the SrO planes [Figs.\ 2(f) and 2(j)]. In particular, a comparison of $dI/dV$ spectra on the right two columns of Fig.\ 2 reveals a clear PG on BiO(II) [Figs.\ 2(g) and 2(o)], although the BiO(I$^\prime$) plane beneath with the same chemical identity exhibits a pronounced DOS enhancement near $E_F$ [Figs.\ 2(h) and 2(p)]. These observations provide compelling evidence that the PG identified on the as-cleaved Bi-2201 is inherent to the BiO plane, echoing our recent experiment on Bi-2212 \cite{lv2015mapping}. Additionally, the PG, no matter on the BiO or SrO planes, develops only after annealing under an appropriate amount of ozone. In combination with the recent demonstration of PG in LaAlO$_3$/SrTiO$_3$ interface \cite{richter2013interface} and potassium-doped Sr$_2$IrO$_4$ with the termination of SrO \cite{kim2014fermi, kim2015observation}, we suggest that PG is inherently a property of electron or hole-doped metal oxides, but not unique to CuO$_2$, although its key mechanism requires further investigations \cite{tahir2011origin, Yamaji2011composite, Feigel2005Superfluid}.

Secondly, the pronounced enhancement in the DOS near $E_F$, which has been interpreted as a van Hove singularity (VHS, or equivalently DOS peak) \cite{lv2015mapping, piriou2011first}, takes place on all the three BiO planes [Figs.\ 2(h), 2(i), 2(k), 2(l) and 2(p)], while the SrO planes are mainly characteristic of a PG after the two cycles of ozone exposure [Figs.\ 2(f) and 2(j)]. This contrasts markedly with Bi-2212, in which VHS is unique to SrO \cite{lv2015mapping}. Such findings indicate that the VHS or primary charge carrier reservoir, which dopes carriers into the CuO$_2$ layers, might take place on distinct ingredient layers: BiO in Bi-2201 and SrO in Bi-2212, respectively. Considering that a VHS emerges only after an ample amount of ozone exposure, we here argue that it should correlate intimately with the interstitial oxygen dopants.

Lastly and most importantly, the PG observed on the BiO(II) plane [Figs.\ 2(g) and 2(o)] appears more well-defined (substantial accumulation of spectral weight at the two gap edges, or stronger coherence peaks) than those observed on BiO(I) [Figs.\ 2(i) and 2(m)] or equivalently BiO-terminated surface of as-cleaved Bi-2201 \cite{Kugler2001Scanning, boyer2007imaging, he2014fermi}. Because of unknown nature of PG, the term `coherence peak' does not have any special physical meaning here. Nevertheless, we use this terminology in order for an easy comparison with previous reports. The PG magnitude $\Delta_p$, measured as half the energy separation between the two coherence peaks, are around 24 meV and 40 meV in Figs.\ 2(g) and 2(o), respectively. Both values appear significantly larger than the values of 12 $\sim$ 16 meV previously reported in Bi-2201 \cite{Kugler2001Scanning, boyer2007imaging, he2014fermi}, which is further confirmed by the line-cut $dI/dV$ spectra with $\Delta_p$ ranging from 28 meV to 49 meV [Fig.\ 3(a)]. The well-defined PGs with pronounced coherence peaks and the so-called `dip-hump' structure [Fig.\ 3(a)] bear great similarities with those previously measured in Bi-2212 \cite{Renner1998pseudogap} and Bi$_2$Sr$_2$Ca$_2$Cu$_3$O$_{10+\delta}$ (Bi-2223) cuprate superconductors \cite{kugler2006scanning}. Historically, $\Delta_p$ has long been argued to correlate closely with the number $n$ of CuO$_2$ planes per unit cell or $T_\textrm{c, max}$: both $\Delta_p$ and $T_\textrm{c, max}$ increase proportionally with $n$ \cite{fischer2007scanning}. However, our finding of the well-defined PG with a comparatively large $\Delta_p$ on the BiO(II) plane in Bi-2201 provides unambiguous evidence that both the $\Delta_p$ and PG have little to do with high-$T_\textrm{c}$ superconductivity in cuprate superconductors. An in-depth understanding of why PG appears so huge and well-defined on the BiO(II) plane of Bi-2201 might turn out to be an important key to unveil the mechanism of PG in cuprates.

Further insights into the PG and VHS are acquired by a spectroscopic study of the as-cleaved Pb-doped Bi-2201 surface. The spatially universal VHS is immediately visible, as illustrated in Fig.\ 3(b). A subsequent UHV annealing gradually removes the excess interstitial oxygen dopants and then VHS. Accordingly, the PG develops near $E_F$ and increases in its magnitude $\Delta_p$ until the spectrum gets featureless after a long-term reduction annealing in UHV. This establishes an intimate relationship between the PG/VHS and excess oxygen dopants in cuprate superconductors, in consistent with the above argument. As compared to PG, the occurrence of VHS requires more interstitial oxygen. In other words, the constituent plane on which VHS is observable must accommodate more excess of oxygen dopants or charge carriers, namely BiO and SrO planes in Bi-2201 and Bi-2212, respectively.

\begin{figure}[t]
\includegraphics[width=\columnwidth]{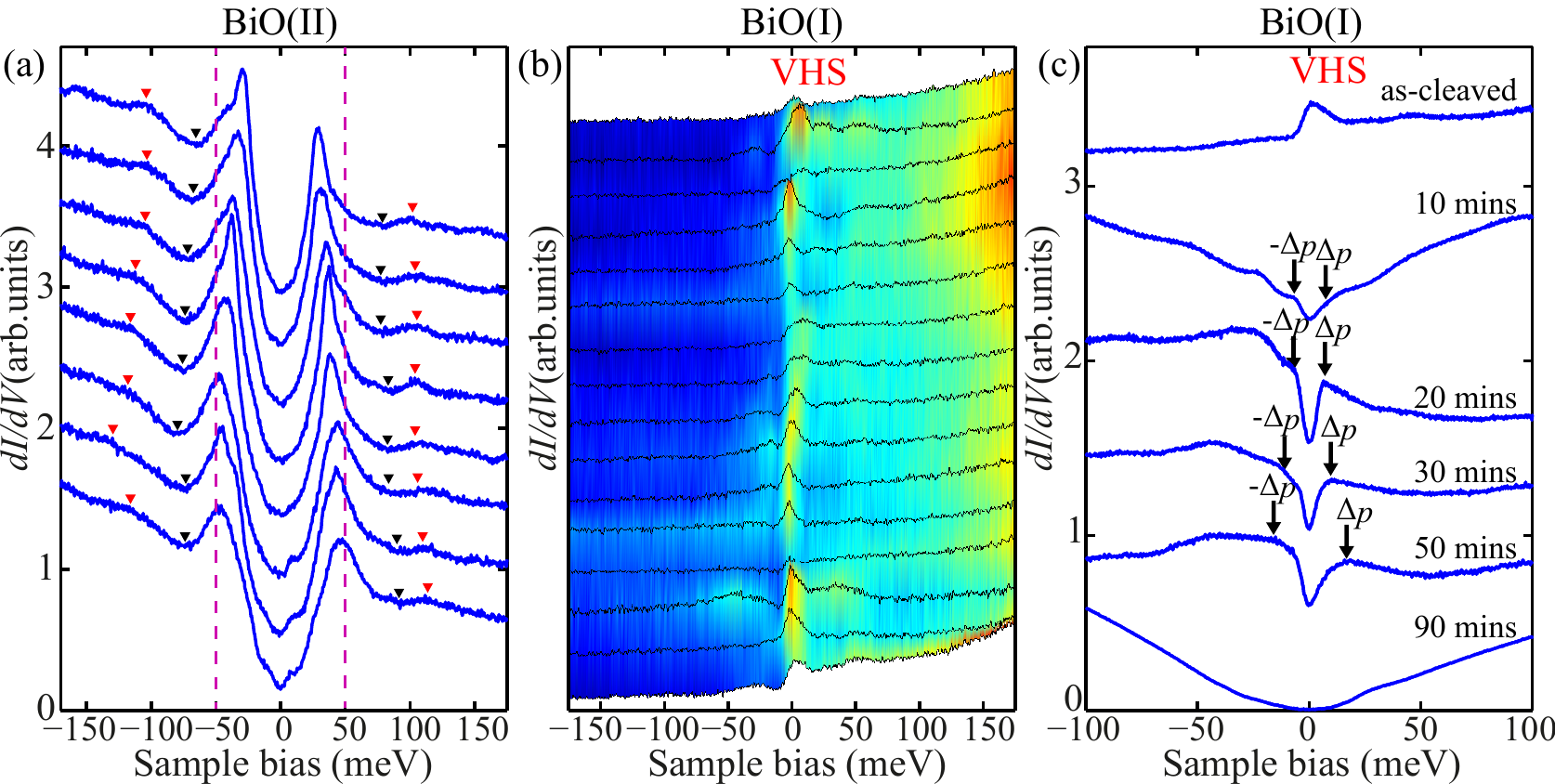}
\caption{(color online) (a, b) Spatially resolved STM tunneling characteristics measured along (a) 8-nm trace on the BiO(II) plane and (b) 30-nm trace on the BiO(I)-terminated surface of as-cleaved Bi-2201, respectively. Black and red triangles denote the dip-hump feature. The spectra have been vertically offset for clarity. (c) UHV reduction annealing dependence of spatially-averaged $dI/dV$ spectra on the as-cleaved BiO(I) plane. The UHV annealing was performed at 500$^\textrm{o}$C, with the respective duration indicated. Setpoint: (a) $V$ = 0.2 V, $I$ = 250 pA; (b, c) $V$ = 0.1 V, $I$ = 200 pA.
}
\end{figure}

Now the natural concerns arise as to why the VHS acting as the primary carrier reservoir occurs distinctively in Bi-2201 and Bi-2212, and how this difference correlates with the different $T_\textrm{c, max}$ in the two cuprate superconductors. To bring insight into these issues, we examine the incommensurate structural buckling in Bi-2212 [Fig.\ 4(a)] and compare with that in Bi-2201 [Fig.\ 1(c)], since such a periodic structural distortion has been consistently revealed to correlate with the excess oxygen distribution \cite{zeljkovic2014nanoscale, Page1989origin, Jakubowicz2001simple}. Generally, a stronger structural buckling means easier and more incorporation of the external oxygen dopants. By comparison of Figs.\ 1(c) and 4(a), we find that the structural buckling behaves conversely in its amplitude between Bi-2201 and Bi-2212. As mentioned above, for example, the structural buckling of BiO plane is stronger than that of SrO in Bi-2201, whereas the opposite holds true in Bi-2212 [Fig.\ 4(a)]. This is more convincingly demonstrated in Fig.\ 4(b), which is based on a statistical analysis of structural buckling amplitude from more than fifty STM topographies. We therefore ascribe the contrasting oxygen or VHS distribution as the different structural buckling in Bi-2201 and Bi-2212 cuprate superconductors.

\begin{figure}[t]
\includegraphics[width=0.81\columnwidth]{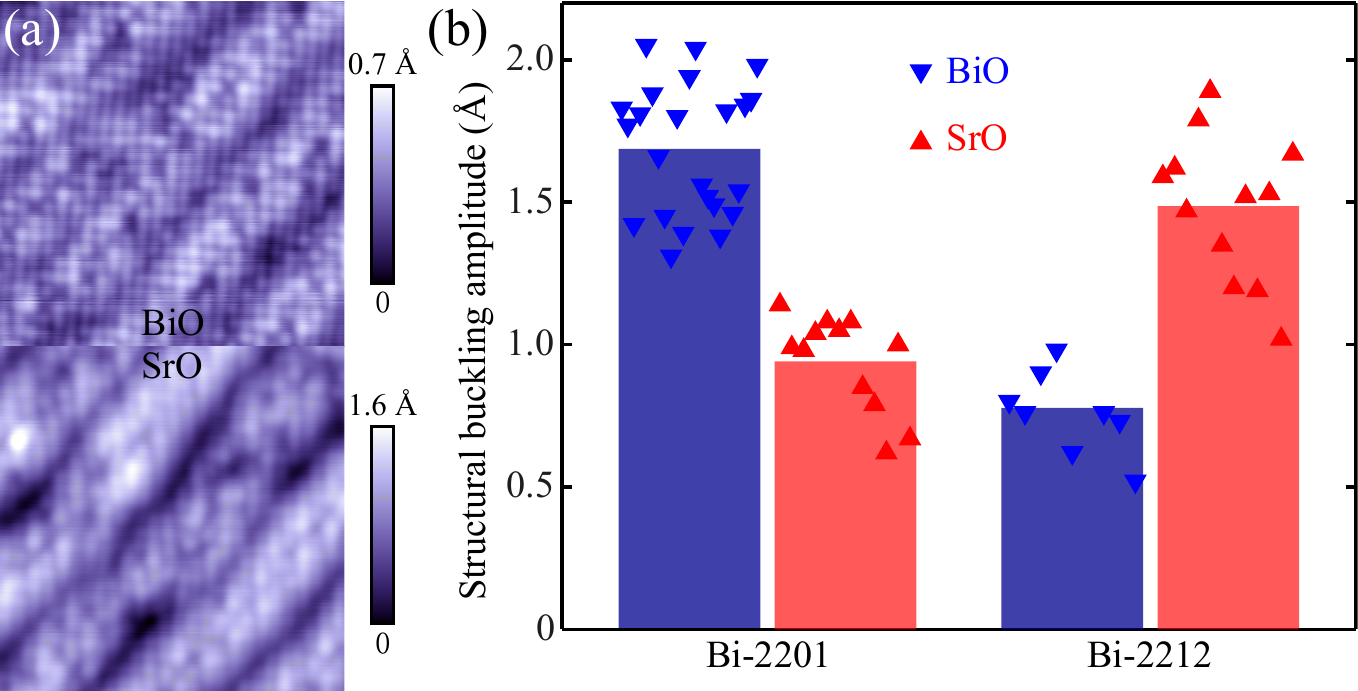}
\caption{(color online) (a) STM topographies ($V$ = 0.2 V, $I$ = 150 pA, 10 nm $\times$ 10 nm) showing the structural buckling of BiO and SrO planes in Bi-2212. (b) Measured structural buckling amplitude on the BiO (blue down triangles) and SrO (red down triangles) planes of Bi-2201 and Bi-2212, namely the maximum height difference of plane subtracted STM images. Each triangle is obtained by estimating the structural buckling amplitude from one STM topography, while the heights of blue- and red-colored bars denote the averaged buckling amplitude for BiO and SrO planes, respectively. Such measurements contain the atomic corrugation, which should be of the same in Bi-2201 and Bi-2212.
}
\end{figure}

Finally we comment on the essential implication of contrasting distribution of VHS or charge carrier reservoir in Bi-2201 and Bi-2212 cuprate superconductors. It is worth noting in Fig.\ 1(a) that the BiO planes are located relatively far away from the major CuO$_2$ planes, with SrO planes in between. As a consequence, the doping efficiency by charge carriers in the BiO planes would be substantially lower than that in SrO. Given that the high-$T_\textrm{c}$ superconductivity develops with carrier doping of CuO$_2$ planes as seen in phase diagram \cite{he2014fermi, fischer2007scanning}, our finding, the charge carriers are predominantly located around the BiO and SrO planes in Bi-2201 and Bi-2212, respectively, accounts excellently for why Bi-2212 has a higher $T_\textrm{c, max}$ than Bi-2201. Indeed, for pure Bi-2201, $T_\textrm{c, max}$ is generally lower than 20 K, whereas La-substituted Bi-2201 (Bi$_2$Sr$_{2-x}$La$_x$CuO$_{6+\delta}$) exhibits a higher $T_\textrm{c, max} >$ 30 K for $x$ $\sim$ 0.4 \cite{kudo2009narrow, lin2010high}. Based on our explanation above, La$^{3+}$ ions substitute for Sr$^{2+}$ sites, acting as positively charged centers, helping attract negatively charged oxygen dopants on the SrO planes of La-substituted Bi-2201. This consequently boosts the superconductivity and leads to higher $T_\textrm{c, max}$ there.

Our detailed STM/STS measurements of the out-of-plane electronic structures have revealed a sharply different charge carrier reservoir in Bi-2201 and Bi-2212 cuprate superconductors. This finding provides a reasonably straightforward explanation why Bi-2212 has a quite higher $T_\textrm{c, max}$ than Bi-2201. Moreover, we discover that the PG exhibits pronounced coherence peaks and significantly enhanced $\Delta_p$ on the BiO(II) plane of Bi-2201, comparable with those reported in Bi-2212 and Bi-2223. Such observation gives definitive proof that PG has little to do with superconductivity in the CuO$_2$ planes. In this respect, our study has provided crucial insights into high-$T_\textrm{c}$ superconductivity in cuprate superconductors.

\begin{acknowledgments}
This work was financially supported by National Science Foundation and Ministry of Science and Technology of China. C. L. S acknowledges support from Tsinghua University Initiative Scientific Research Program. X. J. Z thanks financial support from the MOST of China (973 program: 2015CB921000), the NSFC (11190022 and 11334010) and the Strategic Priority Research Program (B) of CAS with Grant No.\ XDB07020300. Work at Brookhaven was supported by the Office of Basic Energy Sciences (BES), Division of Materials Sciences and Engineering, U. S. Department of Energy (DOE), through Contract No. DE-SC00112704.
\end{acknowledgments}

%

\end{document}